\documentclass[12pt]{article}
\newcommand{\be}{\begin{eqnarray}}
\newcommand{\en}{\end{eqnarray}}

\begin{document}
\begin{titlepage}
\setlength{\textwidth}{5.9in}
\begin{flushright}
EFI 2000-58\\

\end{flushright}
\begin{center}
\vskip 1.0truein
{\Large\bf {Superconvergence and Duality}}
\footnote{For the {\it `Concise Encyclopedia of SUPERSYMMETRY'}, \\
Kluwer Academic Publishers, Dortrecht, (Editors: Jon Bagger, Steven Duplij
and Warren Siegel) 2001.}

\vskip0.4truein
{Reinhard Oehme}
\footnote{E-mail: oehme@theory.uchicago.edu}
\vskip0.4truein
{\it Enrico Fermi Institute and Department of Physics}\\
{\it University of Chicago} \\
{\it Chicago, Illinois, 60637, USA}
\end{center}

\vskip0.7truein

\centerline{\bf Abstract}

\vskip0.2truein
Superconvergence relations for the transverse gauge field
propagator can be used in order to show that the corresponding
gauge quanta are not elements of the physical state space, as
defined by the BRST algebra. With a given gauge group, these
relations are valid for a limited region in the number of
matter fields, indicating a phase transition at the boundary.
In the case of SUSY gauge theories with matter fields in the
fundamental representation, the results predicted by
superconvergence can be compared directly
with those obtained on the basis of duality and the conformal
algebra. There is exact agreement.
\end{titlepage}
\newpage
\baselineskip 18 pt
\pagestyle{plain}
\setlength{\textwidth}{5.9in}

\noindent SUPERCONVERGENCE AND DUALITY.
Superconvergence of the gauge
field propagator leads to general arguments for the absence of transverse
gauge field excitations from the physical state space of the theory. These
arguments are valid for SUSY and for non-SUSY theories. The presence of
superconvergence relations depends upon the gauge group and the the number
and typ of matter fields. Duality is another important concept which
makes statements about the phase structure of SUSY gauge theories. For
SUSY theories of interest, these quite different methods give exactly the
same results.

\noindent It is reasonable to consider $N=1$ SUSY theories
with the gauge group $SU(N_C)$ and $N_F$ massless quark fields in the
fundamental representation. Generalizations will be discussed later.
In the Landau gauge, and for appropriate values of $N_F$, there
exists the SUPERCONVERGENCE RELATION [1]
\begin{eqnarray}
\int_{-0}^{\infty}d k^2 \rho (k^2, \kappa^2, g )
{}~~= ~~0 ~,
\label{1}
\end{eqnarray}
where $\kappa^2<0$ is the normalization point. The integrand
$\rho$ is the discontinuity of the structure function for the transverse
gauge field propagator. It represents the
norm of the states ${\tilde{A}}^{\mu\nu} (k)
\vert 0 \rangle $, where $ A^{\mu\nu} = \partial^{\mu} A^{\nu} -
\partial^{\nu} A^{\nu}$, and $A^{\nu}$ is the component gauge field
of the corresponding superfield. Explicitely,
\begin{eqnarray}
&~&\langle 0 | {\tilde{A}}^{\mu \nu}_{a}(k')
{\tilde{A}}^{\varrho\sigma}_{b}(-k) |0\rangle~~ =~~
\delta_{ab} \theta (k^0) \delta (k'-k) \pi \rho (k^2) \cr
&~& \times (-2){(2\pi)}^4 \left( k^\mu k^\varrho g^{\nu \sigma} -
k^\mu k^\sigma g^{\nu \varrho}
+ k^\nu k^\sigma g^{\mu\varrho} - k^\nu k^\varrho g^{\mu\sigma}\right) ~~.
\label{2}
\end{eqnarray}
With test functions $C^a_{\mu\nu} (k)$, one can form states

$\Psi (C)=\int d^4 k C^a_{\mu\nu} (k) {\tilde{A}}^{\mu\nu}_a
(-k)|0\rangle$,
and obtain the norm
\begin{eqnarray}
\left( \Psi (C), \Psi (C) \right)~ &=&~ \int d^4 k \theta(k^0)
\pi \rho (k^2) C(k) ~~, \cr
C(k)~ &=&~ -8(2\pi)^4 k^{\mu}{\overline{C}}^a_{\mu\nu}(k) k^{\rho}
C^a_{\rho\sigma} g^{\nu\sigma} ~,
\label{3}
\end{eqnarray}
where $C(k) > 0$ for $k^2 \geq 0, ~k^0 \geq 0$.
For positive values of the gauge parameter $\alpha$, there is constant
term $\alpha/{\alpha}_0$ on the righthand side of Eq.(1), where
$\alpha_0 = - {\gamma_{00} }/{\gamma_{01}}$. Here $\gamma(g^2,\alpha)
= (\gamma_{00} + \alpha \gamma_{01} ) g^2 + \cdots$ is the anomalous
dimension of the gauge field. But for the problem of phase transitions,
it is sufficient to work in the Landau gauge.

\noindent For values of $N_F$ and $N_C$ for which the superconvergence
relation is valid, one can show that the transverse gauge field quanta must
be confined in the sense that they are not elements of the physical state
space of the theory. The physical state space is defined in terms of a BRST
cohomology, and is invariant under Poincar\'{e} and equivalence
transformations.
In the space with indefinite metric, it requires a detailed analysis of the
states
${\tilde{A}}^{\mu\nu} (k)\vert 0 \rangle$ in order to proof this fact.
One uses invariant projections into the subspace of positive energy states,
and proceeds to isolate the subset of physical states. A complete
discussion may be found in [2]. There are no superconvergence relations
for the quark propagator. But for the typ of theories considered, the
absence of
all gluon states from the physical subspace should also imply the absence
of quark states associated with the same gauge group [2,3].

\noindent The values of $N_F$, for which the superconvergence
relation (1) is valid, are obtained from the leading
asymptotic term of the gauge field structure function $D(k^2, \kappa^2, g)$
for $k^2 \rightarrow 0$ in all directions of the complex plane.
This term is given by
\begin{eqnarray}
-k^2 D(k^2,\kappa^2,g) & \simeq&
C(g^2) \left(-\beta_0 \ln \frac{k^2}
{\kappa^2}\right)^{-\gamma_{00}/\beta_0} + \cdots ~.
\end{eqnarray}

\noindent The essential aspect of Eq.(4) is the exponent
$({-\gamma_{00}/\beta_0})$,
where $\beta_0$ is the coefficient in the $\beta$-function
$\beta(g^2)= \beta_0 g^4 + \cdots$, and $\gamma_{00}$ has been defined
before. (This exponent is actually the same for all gauges $\alpha \geq 0$).
It is assumed that $\beta_0 < 0$
corresponding to asymptotic freedom. Under these
circumstances, the superconvergence relation (1) is valid
for $\gamma_{00}/\beta_0~>~0$. With the coefficients given by
\begin{eqnarray}
\beta_0~&=&~(16\pi^2)^{-1} (-3N_C ~+~ N_F )\cr
\gamma_{00} ~&=&~(16\pi^2)^{-1} (-\frac{3}{2} N_C ~+~ N_F)~,
\end{eqnarray}
this condition corresponds to $N_F<\frac{3}{2}N_C$ [4]. For values of $N_F$
below the
point $N_F=\frac{3}{2}N_C$, there is confinement of the gauge quanta
associated with the group $SU(N_C)$ as described above. Above this point,
there is the interval
\begin{eqnarray}
\frac{3}{2}N_C~<~N_F~<~3N_C~ ,
\end{eqnarray}
where the theory still has asymptotic freedom, but there is no
constraint on transverse gauge excitations. A detailed study of the
dispersion representations for the propagator and for projected
propagators indicates that no confinement is expected in this region [2].

\noindent The results described above show that the zero point of
the anomalous dimension coefficient $\gamma_{00}(N_F)$ is as important
for the phase structure of the theory as is that of the coefficient
$\beta_0(N_F)$. Both coefficients are invariant under equivalence
transformations
of the theory, and hence scheme independent. Although fields without mass
terms are considered here, if masses of matter fields are present, one can
arrange for mass independent coefficients.

\noindent The consequences of superconvergence,
which are reported above for $N=1$
supersymetric theories, were first obtained for non-SUSY theories
in [2], and then for the SUSY case in [4], where the phase change at
$N_F = 3/2N_C$ is pointed out explicitely. In the $SU(N_C)$
gauge theory without supersymmetry, the interval corresponding
to Eq. (6) is given by
\begin{eqnarray}
\frac{13}{4}N_C~<~N_F~<~\frac{22}{4}N_C~ .
\end{eqnarray}
For $N_F<\frac{13}{4}N_C$, the superconvergence relation is
valid, and hence transverse gauge field quanta are
confined in the sense that they are not in the physical state space.
One may expect that the known perturbative zero
${g_0}^2$ of the renormalization group function $\beta(g^2)$,
at the upper end of the interval (6) or (7) respectively,
actually corresponds to a
general non-pertubative infrared fixed point in the corresponding
window [5].

It is very interesting to compare the phase structure
obtained from superconvergence with the results following from
electric-magnetic DUALITY of $N=1$ SUSY gauge theories [6]. As above, the
electric theory is considered to have the
gauge group $G=SU(N_C)$ and massles matter fields in the representation
$N_F \times ({\bf N_C} + {\bf {\overline {N}}_C})$. It is proposed that
there exists a dual magnetic theory which provides an equivalent description
in the infrared. For appropriate values of $N_C$ and $N_F$, it has the gauge
group
$G^d=SU(N^d_C)$ with $N^d_C=N_F-N_C$.
There are $N^d_F=N_F$ quark
superfields $q$ in the fundamental representation, the corresponding
anti-quark superfields $\overline{q}$, and $N_F^2$ independent
scalar superfields $M$, which are coupled via a Yukawa superpotential of the
form
$\sqrt{\lambda}M^i_jq_i{\overline{q}}^j$.
This coupling is required by the anomaly matching conditions, which are
used in the construction of dual theories. In the conformal window,
the potential drives the magnetic theory to a fixed point which is the same
as the one for the electric theory.

In order to have a dual magnetic theory with a single coupling
parameter, one can use the method of the REDUCTION OF COUPLINGS,
which is a consequence of the renormalization group equations [7].
For the dual magnetic theory discussed here, the reduction equations
have a unique solution with an asymptotic power series expansion.
This solution expresses the Yukawa coupling $\lambda$ in terms of the
magnetic gauge coupling: $\lambda({g_d}^2)=f_0(N_F,N_C){g_d}^2$, where the
coefficient $f_0$ is give explicitly.

\noindent In the reduced, as well as in the two-coupling form
of the magnetic theory, there is no contribution from the
superpotential to the
one-loop coefficients $\beta^d_0$ of the $\beta^d$ function and
$\gamma^d_{00}$ of the anomalous dimension $\gamma^d$ for the
magnetic gauge field. These coefficients are given by
\begin{eqnarray}
\beta^d_{0} ~&=&~(16\pi^2)^{-1} (-2N_F ~+~ 3N_C) \cr
\gamma^d_{00} ~&=&~(16\pi^2)^{-1} (-\frac{1}{2} N_F ~+~ \frac{3}{2}
N_C ) ~,
\end{eqnarray}
where $N^d_F=N_F$ has been used in order to evaluate both theories at the
same number of flavors. However, here these flavors refer to the magnetic
gauge group $G^d=SU(N^d_C)$.

\noindent Comparing the coefficients for the magnetic theory with those for
the corresponding electric theory as given in eq.(5), there emerge
the DUALITY RELATIONS [8]

\begin{eqnarray}
- 2 \gamma_{00} (N_F) ~&=&~ \beta^d_0 (N_F) ~,\cr
- 2 \gamma^d_{00} (N_F)~&=&~\beta_0 (N_F) ~,
\end{eqnarray}

\noindent with $N_F$ on both sides again referring to the different
gauge groups. The factor two is due to the definition of the
anomalous dimension used. These duality relations explicitly
verify the result obtained from superconvergence, namely
that the anomalous dimension coefficient $\gamma_{00}(N_F)$ is of direct
importance for the phase structure. With the
opposite sign, it exactly coincides with the
$\beta$ function coefficient of the dual magnetic theory.
The superconvergence arguments imply that,
as a consequence of the change of sign of $\gamma_{00} (N_F)$,
the electric theory is in a different phase for
$N_F<\frac{3}{2}N_C$ than inside the window (6).
With duality, the magnetic description changes from
asymptotic freedom inside the window to infrared freedom for
$N_F<\frac{3}{2}N_C$. Below the zero
of $\beta^d_0 (N_F)$, the excitations associated with the
magnetic gauge group $SU(N_F-N_C)$ describe the spectrum
for $N_F > N_C + 1$, assuming $N_C>4$.
They may be considered as composites of the confined electric
quanta. Near $N_F=3N_C$, the upper end of the window,
the r\^{o}les of electric and magnetic theories are interchanged .
With $\gamma^d_{00}/\beta^d_0~>~0$ and asymptotic freedom,
the superconvergence argument implies that for $N_F>3N_C$
the magnetic gauge
quanta are not in the physical state space of the magnetic theory.
The infrared free electric quanta describe the spectrum
at low energies, assuming some embedding of the electric
theory at small distances,
for instance into a string theory.
It should be mentioned that the lower limit of the conformal
window follows also from the superconformal algebra, given the
presence of the non-trivial fixed point.

\noindent The duality relations (9) can be shown to be valid for SUSY
theories with other gauge groups, and matter fields in the fundamental
representation. For example,
electric theories with the groups $SO(N_C),~Sp(2N_C), ~G_2$ and
the corresponding dual magnetic
theories with $SO(N_F-N_C+4),~Sp(2N_C-2N_C-4),~SU(N_F-4)$
have been considered. The relations (9) are usually not valid for
theories where matter fields in the adjoint representation are present.
In order to have matching anomalies in these
cases, it turns out that superpotentials are needed for the electric as well
as for the magnetic theories [9]. These potentials then dominate the
phase structure, and the gauge
coupling plays a lesser r\^{o}le. Even though the duality
relations (9) may not
hold, the superconvergence arguments themselves are still of interest.
In particular, superconvergence may give some information about
phases of these theories without superpotentials, where there is no duality.

\newpage

\noindent BIBLIOGRAPHY.

[1] R. Oehme, W. Zimmermann, Phys. Rev. {\bf D21} (1980) 471, 1661;
R. Oehme, Phys. Lett. {\bf B252} (1990) 641;
R. Oehme, W. Xu, Phys. Lett. {\bf B333} (1994) 172, hep-th/9406081,
{\bf B384} (1996) 269, hep-th/9604021..

[2] R. Oehme, Phys. Rev. {\bf D42} (1990) 4209,
Phys. Lett. {\bf B195} (1987) 60;
K. Nishijima, Prog.Theor.Phys. {\bf 75} (1986) 1221.

[3] K. Nishijima, Prog.Theor.Phys.{\bf 77} (1987) 1053;
R. Oehme, Phys. Lett. {\bf B232} (1989) 489;
T. Kugo, I. Ojima, Prog.Theor.Phys.Suppl. {\bf 66} (1979) 1.

[4] R. Oehme, `Superconvergence, Supersymmetry
and Conformal Invariance', in
Leite Lopes Festschrift, pp. 443-457, World Scientific 1988,
(KEK Library Scan, HEP SPIRES SLAC: Oehme, R.).


[5] T. Banks, A. Zaks, Nucl. Phys. {\bf B196} (1982) 189;
E. Gardi, G. Grunberg, JHEP {\bf 024} (1999) 9903, hep-th/9810192.

[6] N. Seiberg, Nucl. Phys. {\bf B435} (1995) 129, hep-th/9411149.

[7] R. Oehme, Phys. Rev. {\bf D59} (1999) 105004-1, hep-th/9808054,
`Reduction of Coupling Parameters in Quantum Field Theories', in
Concise Encyclopedia of SUPERSYMMETRY, Kluver 2001, hep-th/0012162,
Prog. Theor. Phys. Suppl. {\bf 86} (1986) 215;
R. Oehme, W. Zimmermann, Commun.Math.Phys. {\bf 97} (1985) 569;
R. Oehme, K. Sibold and W. Zimmermann, Phys. Lett. {\bf B147} (1984) 115,
Phys. Lett. {\bf B153}(1985) 142;
W. Zimmermann, Commun. Math. Phys. {\bf 97} (1985) 211.

[8] R. Oehme, Phys. Lett. {\bf B399} (1997) 67, hep-th/9701012,
`Superconvergence, Confinement and Duality', Novy Svit, Crimea, 1995:
Proceedings pp. 107-116, Kiev 1995, hep-th/9511014.

[9] D. Kutasov, A. Schwimmer, Phys.Lett. {\bf B354} (1995) 315,
hep-th/9505004;
M. Tachibana, Phys. Rev. {\bf D58} (1998) 045015, hep-th/9709186.

\noindent {\it Reinhard Oehme}

\end{document}